\documentclass[11pt]{article}
\pdfoutput=1
\usepackage[margin=1in]{geometry}
\usepackage{amsmath, amssymb, braket, dsfont,authblk}
\usepackage{graphicx}
\usepackage{rotating, mathtools}
\usepackage[matrix,arrow]{xy}
\usepackage[numbers]{natbib}
\usepackage[colorlinks]{hyperref}
\hypersetup{linkcolor={blue}}

\newcommand{\ba}{\begin{align}}
\newcommand{\ea}{\end{align}}

\begin{document}

%
%
%
%
\title{Deconfinement transitions in a generalised XY model}%

\author{Pablo Serna$^1$, J.T.\ Chalker$^2$, and Paul Fendley$^{2,3}$ 
\medskip \\ 
\small{
$^1$ Laboratoire de physique statistique, D\'epartement de physique de l'ENS, \'Ecole normale sup\'erieure,\\ \emph{and} PSL Research University, Universit\'e Paris Diderot, Sorbonne Paris Cit\'e, \\ \emph{and} Sorbonne Universit\'es, UPMC Univ. Paris 06, CNRS, 75005 Paris, France,\\
$^2$ Rudolf Peierls Centre for Theoretical Physics, University of Oxford,
1 Keble Road, Oxford OX1 3NP, UK\\ 
$^3$ All Souls College, Oxford OX1 4AL, UK }}
\smallskip 

\maketitle

\begin{abstract}
We find the complete phase diagram of a generalised XY model that includes half-vortices.
The model possesses superfluid, pair-superfluid and disordered phases, separated by Kosterlitz-Thouless (KT) transitions for both the half-vortices and ordinary vortices, as well as an Ising-type transition. There also occurs an unusual deconfining phase transition, where the disordered to superfluid transition is of Ising rather than KT type. We show by analytical arguments and extensive numerical simulations that there is a point in the phase diagram where the KT transition line meets the deconfining Ising phase transition. We find that the latter extends into the disordered phase not as a phase transition, but rather solely as a deconfinement transition. It is best understood in the dual height model, where on one side of the transition height steps are bound into pairs while on the other they are unbound. We also extend the phase diagram of the dual model, finding both $O(2)$ loop model and antiferromagnetic Ising transitions.

\end{abstract}

\section{Introduction}

Many physically important theories exhibit confinement, where certain microscopic excitations of the system occur in the long-distance limit only as part of some bound state. Much interesting behaviour occurs at deconfinement transitions, where the excitations become unbound. One famous example occurs in lattice gauge theory, where deconfinement can happen smoothly, with no sharp phase transition \cite{Fradkin:1978}.  Another is the ``deconfined quantum criticality'' scenario \cite{Senthil}. Here different phases are characterised by order parameters coming from distinct symmetries, so that Landau theory predicts a first-order transition between them. Nonetheless it is argued that both order parameters can be expressed in terms of underlying fractionalised quasiparticles, and that critical fluctuations can suppress the binding. If so, the transition is a critical deconfinement transition.

Models argued to have deconfined criticality are typically two-dimensional quantum models, with the physics closely related to the fractionalised quasiparticles that can occur. Our purpose here is to describe in depth a two-dimensional classical model with much of the same physics but more amenable to reliable analysis. This is a venerable model of statistical mechanics, originally introduced to understand the physics of the binding and unbinding of ``half-vortices'' \cite{Korshunov,Lee:1985}. The canonical model of this type is an XY model generalised to include another term with $\pi$ periodicity. The degrees of freedom are rotors $ 0\le \theta_j<2\pi$ at each site $j$ of the square lattice, and all nearest neighbours interact with energy
\begin{align}
E= - \sum_{<jk>} \left[(1-\Delta) \cos(\theta_j-\theta_k) + \Delta \cos(2\theta_j-2\theta_k)\right]\  .
\label{Exy}
\end{align}
The usual XY model has $\Delta=0$, while $\Delta=1$ yields an XY model with half the periodicity (we can identify $\theta_j\sim\theta_j+\pi$).  In an associated quantum chain, the two terms in (\ref{Exy}) correspond respectively to the hopping of single bosons and of boson pairs.

The physics in the XY model at  $\Delta=0$ is very well understood \cite{Chaikin00}. At low temperatures, the partition function is dominated by configurations where $\theta_j$ varies slowly in space. One can then describe it effectively by a field $\theta(\vec{x})$ with energy quadratic in derivatives, i.e.\ a free massless boson. Because this field can describe the phase of the superfluid part of the wavefunction in a helium film, this low-temperature phase  is typically called the superfluid phase. The fact that $\theta\sim\theta+2\pi$ gives rise to vortices and anti-vortices, configurations where around a circle in space, $\theta_j$ or $\theta(x)$ winds gradually from $0$ to $2\pi$. The field varies rapidly only near the vortex center, and so one can think of these as localised excitations. Standard calculations show that the energy of a vortex/anti-vortex pair grows with their separation, so at low enough temperatures, these pairs are bound. In the free-boson field theory, the operator creating a vortex/anti-vortex pair is irrelevant at low temperature. However, as the vortex and anti-vortex get farther apart, more configurations are possible. Thus as the temperature is increased, the vortex/anti-vortex attraction is overcome by the ensuing increased entropy in the free energy, and the famed Kosterlitz-Thouless (KT) transition occurs at $T=T_{KT}=0.893\dots$ \cite{Hasenbusch05}. In the field theory, the pair-creation operator becomes relevant at the KT transition. Thus for higher temperatures, vortices and antivortices are unbound, and the physics is no longer described by a slowly varying field $\theta$.

At $\Delta=1$ the physics is similar. Here there occur half-vortices, configurations where $\theta$ winds gradually from $0$ to $\pi$. In the associated quantum chain, the low-temperature phase can be characterised as a superfluid of boson pairs, so we call this the ``pair-superfluid'' phase.  One way of destroying the pair-superfluid phase is by the unbinding of half-vortex/half-anti-vortex pairs. Since this pair's energy is a fourth of that of a vortex/anti-vortex pair, the dimension of the former pair-creation operator is a fourth of the latter. The ensuing ``half-KT'' transition to a disordered phase therefore occurs at approximately $T_{KT}/4$. 

Varying $\Delta$ away from these special values yields a rich phase diagram, the topic of this paper. In addition to half-vortices, domain walls are important excitations for $\Delta$ slightly smaller than one. These are introduced into an initially smooth spin configuration by making the change $\theta_j \to \theta_j + \pi$ within the region enclosed by a domain wall. Domain walls cost an energy per unit length that is positive for $\Delta<1$ and vanishes as $\Delta \to1$. Including fluctuations, their free energy per unit length decreases with increasing temperature and vanishes at an Ising transition between the superfluid and pair-superfluid phases \cite{Korshunov,Lee:1985}. 

A key issue we address is {\em confinement}: whether a pair of half-vortices must be bound into a vortex, or are free to wander separately. The Ising transition is a deconfinement transition for half-vortices, as may be seen by considering a system with fixed-spin boundary conditions that impose a winding of $\theta_j$ by $2\pi$ around the boundary. Vortex energetics alone would suggest that low-energy configurations with these boundary conditions contain two half-vortices at a separation proportional to system size. However, because a domain wall must run between the half-vortices, they are in fact deconfined in this way only in the pair-superfluid phase where the domain-wall tension vanishes.


The model thus has at least three phases (superfluid, pair-superfluid, and disordered), and at least three lines of continuous transitions (KT, half-KT, and Ising).  One important question then is how this all fits together. The early theoretical analysis, reviewed below, indicates that the three transition lines do all enter the same coupling region, and this is confirmed by numerical analysis \cite{Carpenter}. The simplest scenario is therefore that a multicritical point occurs where all three phases and three transition lines meet. 
This scenario is consistent with the original theoretical analysis and the numerics, but one might worry that the true story might be less interesting. The issue is that there are two different order parameters present: the superfluid order parameter, and the Ising one governing the transition between the different superfluid phases. Although the two order parameters are intimately related, no obvious symmetry requires that they have transitions at the same place. Thus Landau theory suggests that a first-order transition occurs at temperatures at the same scale as the couplings, where perturbative arguments are less reliable. 

The good news is that the true story is even more interesting. All three phases meet at a point, but nowhere do all three critical lines meet. Rather the Ising line extends beyond the pair-superfluid boundary, also governing a direct transition from the superfluid phase to the disordered phase \cite{Shi:2011}.  This is certainly unexpected from a Landau point of view, because two symmetry-unrelated order parameters then have a coincident transition along this line of couplings. What happens is that when the Ising degrees of freedom become critical and have large fluctuations,  they suppress the binding of half-vortices. Past the Ising critical line though, this suppression no longer occurs and so both the superfluidity and Ising criticality are destroyed in tandem. This is quite akin to the deconfined quantum criticality scenario \cite{Senthil}.

These previous results open up several new questions. It is clear from \cite{Shi:2011} that the half-KT line terminates in the Ising line at the point where the three phases meet, but it is then unclear how the KT line fits into the story. A second issue is what happens to the Ising transition within the disordered phase: since the Ising degrees of freedom are still present, one might wonder if they allow for any sort of transition. 

In this paper we address both questions and show that their answers are closely related. We study the phase diagram of a model with the same physics as (\ref{Exy}), and show via extensive numerics that the KT line meets the extended Ising line, with no evidence of a first-order transition. Moreover, we show that at still higher temperatures, the Ising line then heads off into the disordered phase not as a phase transition, but rather as a {\em deconfinement transition}. Roughly speaking, this deconfinement transition amounts to allowing vortices to split into two half-vortices, but since it is within the disordered phase where vortices proliferate, it is hard to characterise it precisely in this language. We therefore study it within a dual model where it can be defined and studied precisely, as the deconfinement of doubled domain walls into individual ones.

\section{The generalised XY model}

In this section we review the construction of a dual model where the degrees of freedom are integer-valued heights on the dual lattice. We follow the earlier analyses by modifying the precise form of the energy in a way that gives a model easier to handle analytically and also makes the Ising structure present much more transparent.
The trick is a well known one due to Villain. For the XY model, one replaces the Boltzmann weight for each nearest-neighbour by 
\begin{align}
e^{\frac{1}{T}\cos(\theta)} \rightarrow \sum_{p=-\infty}^{\infty} e^{-\frac{J}{2}(\theta-2\pi p)^2} \equiv w(\theta)\ .\end{align}
The right-hand-side is periodic under $\theta\to\theta+2\pi$, and is sharply peaked at $\theta=0$ for $J$ large, as the XY Boltzmann weight is. The physics is thus expected to be the same.
The advantage of the Villain potential is that it has a elegant dual form
\begin{align}
w(\theta) = \frac{1}{\sqrt{2\pi J}} \sum_{n=-\infty}^{\infty} e^{in\theta} e^{-\frac{1}{2J}n^2}
\label{wduality}
\end{align}
found by using the Poisson summation formula.

In the generalised XY model for $1/5<\Delta\le 1$, the energy (\ref{Exy}) for adjacent sites has minima at both $\theta_j-\theta_k=0$ mod $2\pi$ and $\pi$ mod $2\pi$. Here there occur half-vortices where the field $\theta$ winds around by $\pi$.  In this region a Villain-style replacement capturing this physics is \cite{Korshunov,Lee:1985}  
\begin{align}e^{\frac{1}{T}[(1-\Delta)\cos(\theta)+\Delta\cos(2\theta)]} \rightarrow  w(\theta)+e^{-K}w(\theta+\pi)\ , 
\end{align}
where the coupling $K$ controls the weighting of half-vortices.
The partition function is then
\begin{align}
Z= \prod_{j}\int_{-\pi}^\pi d\theta_{j}\, \prod_{<jk>} \left(w(\theta_{jk})+e^{-K}w(\theta_{jk}+\pi)\right)\ 
\label{ZVillain}
\end{align}
where $\theta_{jk}\equiv \theta_j-\theta_k$.
The model (\ref{ZVillain}) exhibits the key features of the generalised XY model. 
The minima at $\theta=0$ and $\pi$ are degenerate at $\Delta=1$ in the original model (\ref{Exy}), so there is no impediment to a vortex splitting into two half-vortices. The same degeneracy occurs at $K=0$, and indeed here (\ref{ZVillain}) reduces to the Villain form of the XY model with periodicity $\theta\sim\theta+\pi$. Reducing $\Delta<1$ or analogously making $K$ positive gives $\theta=\pi$ a higher energy and thus a smaller Boltzmann weight. At $K\to\infty$ we recover the original XY model.

A key question we address in this paper is whether half-vortices are confined, i.e.\ whether the dominant contributions to the free energy are configurations where half-vortices only appear close to either a half-anti-vortex, or another half-vortex.   The potential winds by $2\pi$ around a circle enclosing the two half-vortices, and when they are confined they can effectively be treated as a single vortex. 
Since the potential has a minimum at $\theta_{jk}=\pi$, it is energetically favorable for the field to jump quickly by $\pi$ along a line emanating from a half-vortex.  This line must then terminate in an anti-half-vortex or another half-vortex.  It is conventional to say that the two are then ``attached by a string''. Since $w(0)>w(\pi)$, for $K$ finite and positive $w(0)+e^{-K}w(\pi)>w(\pi)+e^{-K}w(0)$. Thus the potential minimum at $\theta_{jk}=\pi$ has a higher energy than that at $\theta_{jk}=0$, and there is a ``string tension'', an energy cost growing with the length of the string. However, the longer the string, the more possible configurations there are, and so their entropy increases. Thus understanding the deconfinement transition is related to understanding the statistical mechanics of the strings. 

A beautiful and precise way to disentangle the physics of this string from the usual KT physics comes from exploiting the duality of the Villain weight \cite{Korshunov,Lee:1985}. Using the dual form (\ref{wduality}) in (\ref{ZVillain}) means that the $\theta_j$ integrals are easily done:
\begin{align}
\nonumber
Z&\propto
\prod_{j}\int_{-\pi}^\pi d\theta_{j} \prod_{<jk>}\,
\sum_{n_{jk}=-\infty}^{\infty}e^{in_{jk}\theta_{jk}} e^{-\frac{1}{2J}n_{jk}^2}\left(1+e^{-K}(-1)^{n_{jk}}\right)\\
&\propto \sum_{\{n_{jk}\}|\nabla\cdot n=0} e^{-\frac{1}{2J}n_{jk}^2}\left(1+e^{-K}(-1)^{n_{jk}}\right)
\ .
\end{align}
Thus the duality replaces the rotor degrees of freedom with the integer-valued variables $n_{jk}$ that live on every link $jk$ of the lattice. These obey the constraint of vanishing lattice divergence at each site, i.e.\ $\nabla\cdot n \equiv \sum_{k\ {\rm next}\ \rm{to}\ j} n_{jk}=0$ for all sites $j$. Heuristically, these degrees of freedom correspond to currents describing the world lines of bosons, which are conserved because of the vanishing divergence. We emphasise that the mapping to the dual is exact, with no approximations made.

The partition function can be written in a more transparent form in terms of the dual coupling $K_*$ defined by
\begin{align}
e^{K_*}=\coth(K/2)\quad \Leftrightarrow\quad e^{K}=\coth(K_*/2)\quad \Rightarrow\quad \sinh(K)\sinh(K_*)=1
\label{Kstar}
\end{align}
so that
\begin{align}1+e^{-K}(-1)^{n} = \frac{1}{\cosh(K_*/2)}\Big(\cosh(K_*/2)+(-1)^{n}\sinh(K_*/2)\Big) =  \frac{e^{K_* (-1)^{n}/2}}{\cosh(K_*/2)} \ .\end{align}
Using the dual coupling allows the $K$ dependence to be put into the exponent:
\begin{align}
Z
\propto \sum_{\{n_{jk}\}|\nabla\cdot n=0} \exp\left[{-\frac{1}{2J}n_{jk}^2+\frac{K_*}{2}(-1)^{n_{jk}}}\right]\ .
\label{Zn}
\end{align}
The vanishing lattice divergence enables $n_{jk}$ to be rewritten in terms of height variables $h_a$ defined at the sites $a$ of the dual lattice. Each link $jk$ of the original lattice is in one-to-one correspondence with a link $ab$ of the dual lattice. To orient the latter, we split the original lattice into two interpenetrating square sublattices, each with half the sites (e.g.\ the white and black squares on a chessboard), and orient the links $ab$ such that going from $a$ to $b$ is clockwise around points on one such sublattice and counter-clockwise on the other. The integer-valued heights are then
\begin{align}
h_{a}-h_{b} = n_{jk}\ .
\label{hn}
\end{align}
The orientation and the vanishing lattice divergence guarantee that $h_a$ is defined consistently up to an unimportant overall shift. The sum over the $n_{jk}$ can then be replaced by a completely unconstrained sum over heights, giving the partition function used for the rest of the paper:
\begin{align} 
Z \propto \sum_{\{h_a\}}
\exp\left[-\frac{J_*}{2}(h_a-h_b)^2+\frac{K_*}{2}(-1)^{h_a}(-1)^{h_b}\right]\ .
\label{Zh}
\end{align}
Since $J$ appears in the denominator in the exponent, we have defined the dual coupling $J_*=1/J$. Thus the high-temperature phase of the original model corresponds to a weak-coupling phase in terms of $J_*$. 

Note that configurations of the height model for a system with periodic boundary conditions can be separated into sectors classified by their winding number. Specifically, the two net fluxes around the system in a configuration of link currents $\{n_{jk}\}$ are represented in height variables $\{h_a\}$ as integer-valued changes $W_x, W_y$ on encircling the system in each direction. Then in a system of linear size $L$ the height field obeys the boundary conditions $h(x+L,y) = h(x,y) + W_x$ and $h(x,y+L) = h(x,y) + W_y$. The partition function is a sum over height configurations from all sectors but we will also find it useful to analyse behaviour within distinct sectors.

\section{Phases and phase transitions}
\label{sec:phases}

The two terms in the dual height model (\ref{Zh}) describe different physics, and their interplay gives the rich phase diagram of the generalised XY model. The relevance of the Ising physics to the half-vortex story is apparent from the second term of the partition function (\ref{Zh}). We can think of $\mu_a\equiv (-1)^{h_a}=\pm 1$ as an Ising spin, up for even values of the heights, down for odd. The $K_*$ term is then simply the standard Ising 
nearest-neighbour interaction. The three different phases in the generalised XY model and a fourth occurring in the dual model all can be understood by analysing special limits. We do this first, before turning to the more difficult question of what happens in the middle of the phase diagram, where $J_*\sim |K_*|$.

\subsection{Phase transitions of the generalised XY model in extreme limits}

Here we discuss the classic results \cite{Korshunov,Lee:1985} for phase transitions in the generalised XY model in various extreme limits, which are reviewed in \cite{Korshunov06}. 

\paragraph{Ising transition at low temperature/small $J_*$} 
At low temperatures, $J_*=1/J$ is small. Thus in this regime, the latter term in (\ref{Zh}) dominates the energy and the physics is that of the Ising model. 
For $K_*$ large, the Ising spins $\mu_a=(-1)^{h_a}$ order, so the heights are predominantly even or predominantly odd. The $n_{jk}$ defined via (\ref{hn}) are thus predominantly even, and so the bosons in the corresponding quantum chain prefer to form pairs. Large $K_*$ is small $K$ in the original language, so this is the region where half-vortices are allowed.
Thus a precise definition of the pair-superfluid phase is to have a non-vanishing expectation value of $\mu_a$, while still having large fluctuations in $(h_a)^2$. In the original rotor language, it means the half-vortices are bound only with anti-half-vortices, not by confinement into vortices.

Decreasing $K_*$ while still keeping $J_*$ small causes the Ising degrees of freedom to undergo a phase transition at $K_*=K_{\rm Ising}$. At $J_*=0$, $K_{\rm Ising}=\ln(1+\sqrt{2})$, and numerics indicate that the value does not change very much when $J_*$ is increased.  This phase transition is in the Ising universality class, so the correlator of two Ising spins obeys
\begin{align}
\Braket{\mu(\vec{x})\mu(\vec{x'})}\sim |\vec{x}-\vec{x'}|^{-1/4},\qquad\quad\hbox{when }K_*=K_{Ising}.
\label{mumu}
\end{align}
For $|K_*|<K_c$ and $J_*$ small the heights are in a rough phase, where they are completely disordered. In the original rotor language, this is the ordinary superfluid phase. 

\paragraph{KT transition at $K_*$ small}
The effect of increasing $J_*$ or equivalently the temperature while keeping $|K_*|$ small is easy to understand. When $K_*$ is small the Ising term in (\ref{Zh}) can be ignored, and one obtains the height form of the usual XY model with Villain action. The usual Coulomb-gas analysis applies, and the usual KT transition out of the superfluid phase into a disordered phase occurs at $J_*\sim 1.33$.\cite{Janke93} In the height language, (\ref{Zh}) makes it apparent that $J_*$ large favors constant heights. Thus the KT transition for small $K_*$  is from the rough (superfluid) phase into the flat (disordered) phase. 

\paragraph{Half-KT transition at $K_*$ large}
For large positive $K_*$, the Ising interaction favors keeping nearest-neighbor heights both odd or both even, the Ising order of the paired superfluid. Thus for $J_*$ small the heights are still in a rough phase, but with this constraint. Increasing $J_*$ while keeping $K_*$ large still keeps the heights predominantly even or odd. Since here $(h_i-h_j)^2$ is predominantly a multiple of $4$, we effectively can rescale the coupling $J_*\to 4J_*$ and redefine the variables in the corresponding term to be integer. This shows that there is a KT-type transition at $J_*\sim 1.33/4=0.33$. This of course is the half-KT transition from the paired superfluid phase to the disordered phase in the original language. This disordered phase is flat in the height language like that at $K_*$ small, but it is not at all obvious how this flat (disordered) phase is related to the flat (disordered) phase near the KT transition, where there is no constraint that the heights be predominantly even or predominantly odd. This is one of the main questions answered in this paper, and we address it in section \ref{sec:deconfinement}.

\subsection{$K_*$ negative}\label{negK}

The dual model (\ref{Zh}) has the nice feature that it can be extended further in parameter space. The original model (\ref{ZVillain}) is equivalent under sending $K\to -K$, as follows from redefining $\theta_j\to\theta_j+\pi$ on every other site. However, the duality transformation applies only for positive $K$ and $K_*$, as is apparent from the first two maps in (\ref{Kstar}). Thus there is no reason why sending $K_*\to-K_*$ gives an equivalent model, and it is apparent from (\ref{Zn}) or (\ref{Zh}) that it does not. Here we analyse $K_*$ negative, as the methods provide a nice warmup to understanding the subtler deconfinement transition.

\paragraph{The KT transition at $K_*$ negative.}

Having $K_*=0$ corresponds in the original language to sending $K\to\infty$ and recovering the Villain form of the XY model. Increasing $K_*$ slightly does not change the transition type, but merely modifies the critical value $J_c$. In the language of the renormalisation group, including $K_*$ is an irrelevant perturbation. Thus making $K_*$ slightly negative has the same effect in the generalised model; the KT transition persists. 

As $K_*$ is decreased further, there is another way of looking at the KT transition. We thus consider a different region, where $J_*$ and  $- K_*$ are large. Then the two terms compete: the large $J_*$ term favours a flat phase, but the other term with $K_*\sim - J_*$ favours having $(-1)^{h_j}$ alternate. Precisely, from (\ref{Zh}) it follows that the energy of two identical adjacent heights $h_a=h_b$ is $-K_*/2$, while the energy of having $h_a=h_b \pm 1$ is $(J_*+K_*)/2$. Thus the energy of a segment of domain wall is 
$K_*+J_*/2$. We thus expect a transition roughly when the domain-wall energy becomes zero, i.e. $K_*\approx -J_*/2$.

To make this precise, we note that when $0>K_*>-J_*/2$, a good approximation is to neglect all configurations except those with single domain walls, i.e.\ we constrain heights on adjacent lattice sites to differ by only $0,\pm 1$.  Then each height configuration can be written as an {\em oriented loop} configuration on the dual lattice. The loops separate regions of different spins; if the arrow on the loop is going clockwise, $h_{\rm outer}=h_{\rm inner}+1$, while if counterclockwise, $h_{\rm outer}=h_{\rm inner}-1$. (Here we need not worry about the effect of loops that wrap around a cycle, but below such loops will give us an important tool to probe deconfinement.) Two loops can meet at a point, but are forbidden to cross. Up to a subtlety that will be explained below, the partition function (\ref{Zh}) can then be rewritten as a sum over oriented loop configurations. Since loops do not cross, this is equivalent to a sum over unoriented loops where each loop gets a weighting two. Then for ${\cal L}$ an unoriented loop configuration on the square lattice, we let $N({\cal L})$ be the number of such loops in this configuration, $L({\cal L})$ be the combined length of these loops (i.e.\ the number of links on the dual lattice with domain walls), and $T({\cal L})$ be the number of sites on the dual lattice where four domain walls touch. With an appropriate choice of boundary conditions, the partition function can then be written as a sum over loops as 
\begin{align}
Z\propto \sum_{\cal L} 2^{N({\cal L})} z^{L({\cal L})} t^{T({\cal L})}\ .
\label{Zloop}
\end{align}
where $z=\exp(-K_*-J_*/2)$, and we allow for a weight $t$ for each touching.

\begin{figure}[htb!]
\begin{center}
\includegraphics[width=0.5\linewidth]{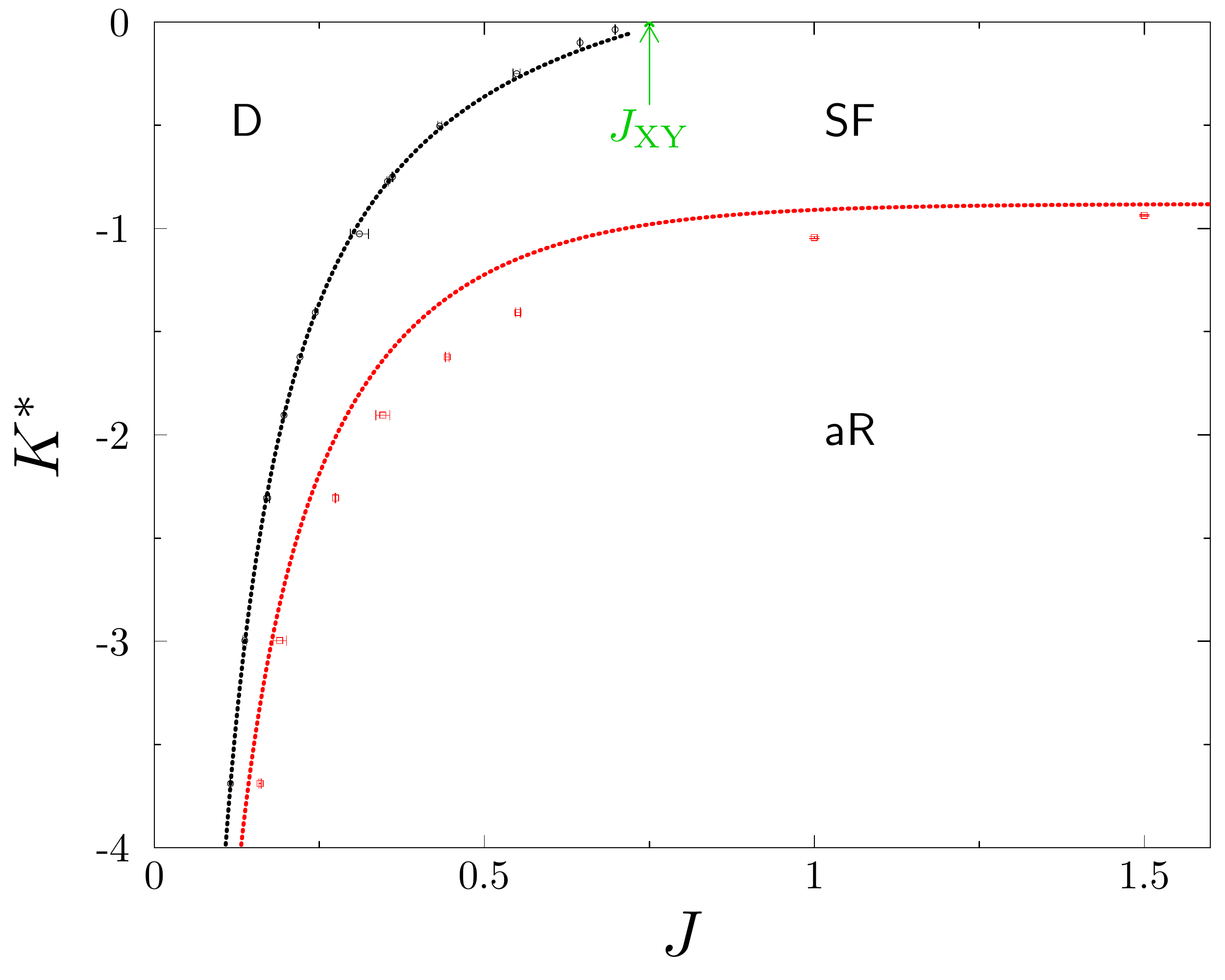}
 \caption{Phase diagram for negative $K_*$ obtained from Monte Carlo simulations,   showing the disordered (D), superfluid (SF) and antiferromagnetic rough (aR) phases and the theoretical predictions for the location of phase boundaries from (\ref{KTcurve}) and (\ref{Isingcurve}), using $z_{\rm KT} = 0.528$ for the former.} 
\label{negKfig}
\end{center}
\end{figure}

The partition function (\ref{Zloop}) is that of the famous dilute $O(2)$ loop model on the square lattice \cite{Blote89}. The reason for the name is that it is in the same universality class as a two-component spin model with $O(2)=U(1)$ symmetry -- the XY model! This XY model is parametrically very far from that at $K_*=0$, the XY limit of (\ref{Zh}).  However, this mapping means that still must be a continuous KT phase transition at a particular value $z=z_c$. Since we already saw that the KT transition extends to $K_*$ small and negative, it is natural to expect that the two lines are one and the same, so that the KT line extends to all $K_*<0$. We thus expect the KT line to be parametrised for all $K_*$ negative as
\begin{align}
K_* \approx -J_*/2 - \ln\left(z_{\rm KT}\right)\ .
\label{KTcurve}
\end{align}
Monte Carlo results (see Sec.~\ref{simulations}) for the phase boundary  are shown in Fig.~\ref{negKfig}.  The data clearly follow the form (\ref{KTcurve}) for all $K_*<0$.

There are several methods of estimating $z_{\rm KT}$, all giving consistent values. 
For $K_*=0$, accurate numerical methods locate the KT transition for the Villain form at $J_{\rm XY}\approx 0.75$,\cite{Janke93,Hasenbusch05}, so this suggests $z_{\rm KT}\approx e^{-0.67}\approx 0.51$.  The loop-model approach suggests a similar value, even though the map to the loop model is not exact,  as it neglects all configurations with anything more than single domain walls (which as we will see play a very important role in the physics once $J_*$ is of order $1$). The parameter $t$ is irrelevant, and while it does not affect the universal properties, it does affect the value of $z_{\rm KT}$. The subtlety alluded to above is that the sum over all ways two oriented loops can touch overcounts the number of configurations in the original height model with single domain walls. A straightforward argument shows that the overcounting can be roughly accounted for by setting $t=2/3$.  The precise location of $z_{\rm  KT}$ for this loop model on the square lattice does not seem to have been studied. Fortunately, the methods of Jacobsen \cite{Jacobsen15} are applicable. He was kind enough to extend his analysis, finding that \cite{Jacobsen17} $z_{\rm KT}(t=0) = 0.59202714$ and $z_{\rm KT}(t=1) = 0.53862569$. Setting $t=2/3$ thus suggests $z_{\rm KT}\approx 0.56$. A final method of finding $z_{\rm KT}$ is simply to fit it from the numerical data in figure \ref{negKfig}. The resulting value $z_{\rm KT}\approx 0.528$ is consistent with the others.


\paragraph{Antiferromagnetic Ising transition}
At $K_*=-K_{\rm Ising}$ and $J_*$ small, the model undergoes another Ising transition, so that for $K_*<-K_{\rm Ising}$, the Ising degrees of freedom are antiferromagnetically ordered. In this ``antiferromagnetic rough'' phase, the heights on the dual lattice alternate predominantly between even and odd, but fluctuate freely otherwise. In this phase, $(-1)^{h_a-h_b}$ for $a$ and $b$ nearest neighbours has a non-vanishing expectation value. This order parameter is invariant under shifting all heights and so remains local in terms of the link variables $n_{jk}$. The antiferromagnetic rough phase is the only phase characterised by such a local order parameter.  No such phase occurs in the original generalised XY model, where $K_*$ must be positive. 

The antiferromagnetic transition persists for all values of $J_*$. Namely, if $K_*\ll -J_*/2$, it follows  from (\ref{Zh}) that the Ising term dominates and yields the antiferromagnetic phase. The location of the transition can be approximated by computing the effective Ising coupling $K_{\rm eff}$ for a single pair of nearest-neighbour heights $h_a$ and $h_b$ which comes from summing the Boltzmann weight in (\ref{Zh}) over all $h_a-h_b$ even and odd respectively. Thus $K_{\rm eff}=\ln(W_{\rm even}/W_{\rm odd})$, where $W_{\rm even}$ and $W_{\rm odd}$ are both written nicely in terms of Jacobi elliptic theta functions:
\begin{align*}
W_{\rm even}&\equiv e^{\frac{K_*}{2}}\sum_{m=-\infty}^\infty e^{-2J_*m^2}\ =\ e^{\frac{K_*}{2}}\vartheta_3(0,e^{-2J_*})\ , \\
W_{\rm odd}&\equiv e^{-\frac{K_*}{2}}\sum_{m=-\infty}^\infty e^{-J_*(2m+1)^2/2}
\ =\ e^{-\frac{K_*}{2}}\vartheta_2(0,e^{-2J_*})\ .
\end{align*}
The antiferromagnetic Ising critical point is at $K_{\rm eff}=-\ln(1+\sqrt{2})$, so the antiferromagnetic Ising critical curve for $K_*$ negative is therefore
\begin{align}
K_* \approx  - \ln(1+\sqrt{2}) + \ln\left(\frac{\vartheta_2(0,e^{-2J_*})}
{\vartheta_3(0,e^{-2J_*})}\right)\ .
\label{Isingcurve}
\end{align}
In the extreme limits the curve becomes approximately 
\begin{align}
K_*\approx 
\begin{cases}
-\frac{1}{2J} - \ln(1+\sqrt{2})+\ln(2)\quad &J \hbox{ small},\\
- \ln(1+\sqrt{2})&J \hbox{ large.}
\end{cases}
\end{align}
Despite coming from an approximation valid at large $J$, the curve at small $J$ has the correct leading term, and a reasonable if not perfect value of the subleading constant.
The antiferromagnetic curve thus never intersects the KT line (\ref{KTcurve}), and so a narrow sliver of the superfluid phase extends to arbitrarily small $J$.
A comparison of these theoretical predictions with the phase diagram for negative $K_*$ determined in simulations is shown in Fig.~\ref{negKfig}.

\section{The deconfinement transition in the disordered/flat phase}
\label{sec:deconfinement}

In the generalised XY model, $J$ small is the high-temperature phase, and there is no impediment to creating vortices and antivortices. These of course disorder the system and destroy the superfluidity. One might expect that disorder means disorder, and that the effect of the Ising term can be ignored for small $J$. This is also natural in the height language, where $J_*$ large strongly favours having the same heights on neighbouring sites and so puts the system in a flat phase. It is not immediately apparent how varying the Ising coupling $K_*$ will affect the flat phase.

One of the central findings we present in this paper is that height steps, occurring in the flat phase at non-zero winding number, undergo a deconfinement transition. The deconfinement transition is not a phase transition in the usual sense. The dominant contributions to the partition function or to the ground state of the corresponding 1d quantum chain do not exhibit any singularities. Instead, on the deconfined side of the transition, individual encircling height steps typically have unit height, while on the confined side they are bound in pairs. 
This deconfinement transition occurs when $J_*$ and $K_*$ are both positive, large and of the same order of magnitude. 
We saw above that in the analogous situation for $K_*$ negative, the competition between the different terms in (\ref{Zh}) results in extensions of the conventional KT transition and antiferromagnetic Ising transitions to strong couplings. Here we show that for $K$ and $K_*$ positive, the corresponding competition results in a subtler but much more interesting transition.



To make this precise, we consider large $J_*$ and $K_*$. As opposed to the $K_*$ negative case treated above, the competition between the two terms in the energy does not affect the dominant state, the flat state where heights are all the same. The interesting interplay occurs in the lowest-lying excited states above the flat state.  It is sufficient to compare the energetics of bound and unbound steps, i.e.\ which type of domain wall separates adjacent heights $h_a$ and $h_b$. From (\ref{Zh}), the energy per unit length of a single-step domain wall with $h_a=h_b\pm 1$ is $J_*/2+K_*$, whereas that of a double-step wall $h_a=h_b\pm 2$ is $2J_*$. Thus
the difference in energy cost per unit length between a double step and two single steps is  $ J_* - 2K_*$. Neglecting entropic contributions we therefore expect two single steps to be unbound if $J_*> 2K_*$ and bound into a double step if $J_* < 2K_*$.  We label these two regions the pair-disordered phase and the disordered phase respectively. This change in behaviour we call the {\em deconfinement transition}.

\begin{figure}[bt!]
\begin{center}
\includegraphics[width=0.5\linewidth]{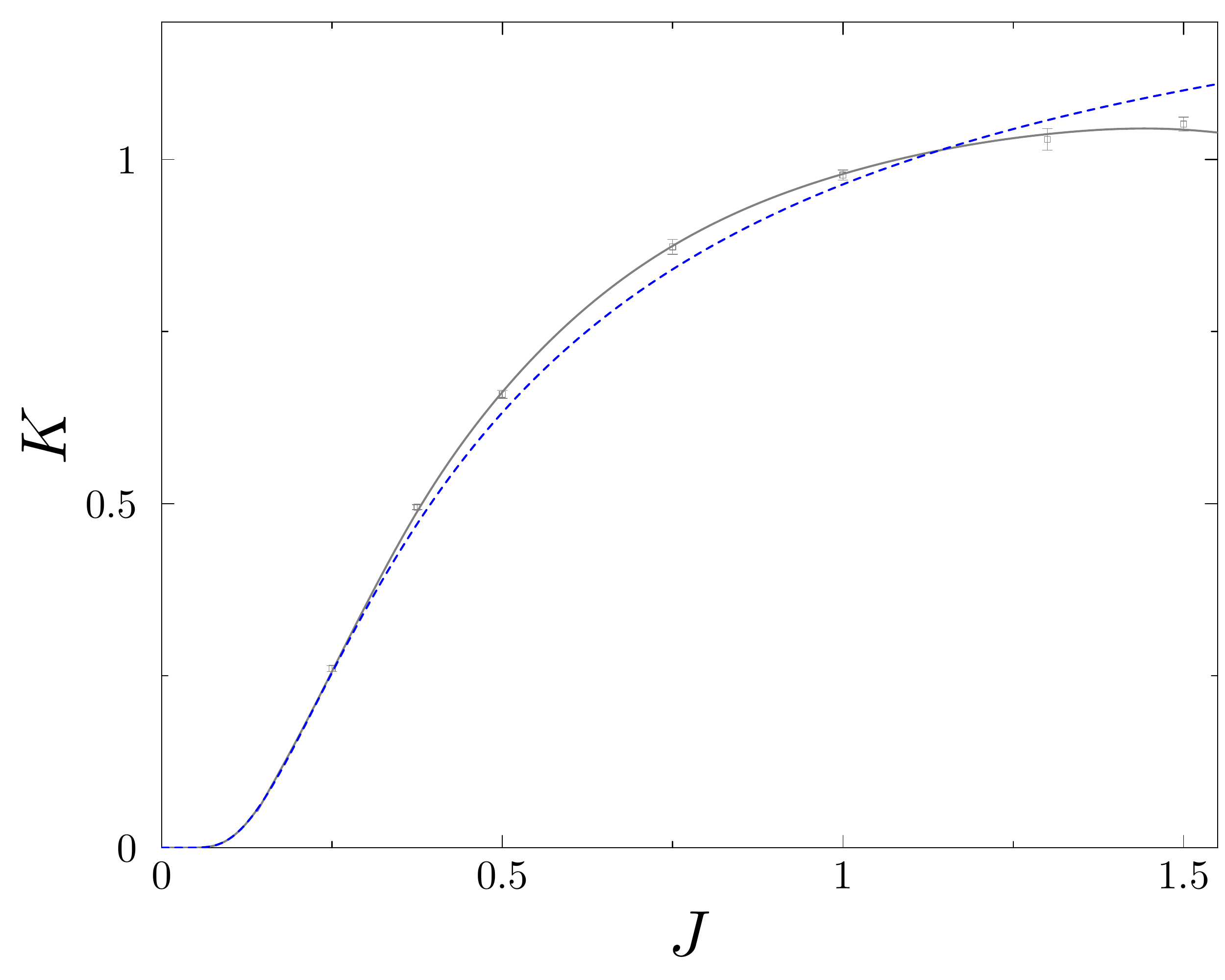}
 \caption{Phase boundary for the deconfinement transition: full line: spline curve drawn though points from Monte Carlo simulations and the origin; dashed line from the asymptotic curve (\ref{deconf1}).}\label{pd-deconf}
\end{center}
\end{figure}

Studying different winding number sectors provides an effective way to probe this deconfinement transition. In quantum language, such a probe gives access to excited states. In the flat phase, the winding number sector $(W_x,W_y)=(0,0)$ contains the flat configuration and all those comprised of short loops. These are the dominant contributions to the partition function, so the probability for the system to be in a sector with non-zero winding number is exponentially small in the linear system size $L$. In a non-zero winding sector, however, we force at least one domain wall to wrap around one cycle. Consider in particular $(W_x,W_y)=(0,2)$. All configurations in this sector contain a net step of two in the height field, encircling the system in the $x$-direction. In the flat phase, other steps in the height field typically consist of short, closed loops and are unimportant for long-distance behaviour.  We thus can ignore the short loops and focus on the behaviour of the net step of two, and so locate the deconfinement transition. The partition function in this winding sector in the flat confined phase is dominated by configurations with one double-step loop wrapping around. In the pair-disordered phase, it is instead dominated by configurations with two single-step loops wrapping around. Using the Monte Carlo numerical methods described in section \ref{simulations}, we locate this deconfinement transition and plot the results in figure \ref{pd-deconf}.

The leading entropic effects can be included at large $J_*$ and $K_*$, giving the leading correction to the deconfinement curve $K_*\approx J_*/2$. In the winding number sector $(W_x,W_y)=(0,2)$, steps in the dominant configurations are more or less straight and run parallel to the ${x}$-direction.  The steps can have dilute kinks but rarely backtrack. Neglecting backtracks, the configuration of the pair of encircling steps is specified by two integer-valued functions $y_l(x)$, $l=1,2$, of an integer-valued coordinate. We are concerned with the step separation $z(x) = y_2(x) - y_1(x)$, which is zero where there is a double step and positive otherwise. Omitting interactions between the encircling steps and short closed-loop excitations, a configurational energy that captures the leading statistical mechanics of the coordinate $z(x)$ is 
\begin{align}
E_{\rm 1D} =  -V \sum_x \delta_{z(x),0} + T \sum_x |z(x+1) - z(x)|
\label{1D}
\end{align}
with $V = 2K_* - J_*$ and $T = J_*/2$. This model is known \cite{Chalker:1981} to have an unbinding transition at $e^{-V}+e^{-T} =1$. The deconfinement transition at large $K_*,J_*$ therefore occurs at
\begin{align}
K_*\approx \frac{J_*}{2} + \frac{e^{-J_*/2}}{2} + O(e^{-J_*})\ .
\label{deconf1}
\end{align}
In terms of the original generalised XY couplings, this translates to
\begin{align}
K \approx 2e^{-1/(2J)} - e^{-1/J} + O(e^{-3/(2J)})\ .
\label{deconf2}
\end{align}
at $K$ and $J$ small.

\section{The full phase diagram}

In the preceding sections we analysed the generalised XY model in various analytically tractable limits. We often used the dual description as a height model, whose flat phase corresponds to the disordered phase of the XY model. We found in particular a previously unexplored deconfinement transition where this phase splits into two: a region where the domain walls are confined into double steps and a region where they are not. The transition between the two occurs along a sharp line, parametrised for large dual couplings as (\ref{deconf1}), and for small original couplings as (\ref{deconf2}).

In this section we show how all these phases fit together in the  ``middle'' of the phase diagram, where both couplings $J_*$ and $K_*$ are order $1$. A particularly interesting question we address is the way they mesh with the  deconfinement {\em phase} transition found in \cite{Shi:2011}. As opposed to the deconfinement transition in the flat phase described above, this deconfinement is a genuine phase transition of Ising type. This critical line describes a  transition from the superfluid phase directly into the pair-disordered phase, a classical two-dimensional analog of the deconfined quantum criticality scenario \cite{Senthil}. Both the superfluidity and the half-vortex confinement have critical phase transitions at the same coupling, even though the transitions seem to arise from distinct order parameters. 

This Ising deconfinement phase transition is straightforward to see in effective field theory \cite{Shi:2011}. The XY rotor gives rise to a bosonic field in the usual fashion, while the Ising degrees of freedom are represented by the disorder field of the Ising conformal field theory. In the field-theory limit of the partition function, the two fields are coupled by the half-vortex creation operator {\em times} a disorder field. Thus the fluctuations of the Ising degrees of freedom can suppress those of the XY boson, because the combined operator can be irrelevant even when the individual operators are relevant. The Ising transition therefore persists for smaller $J$ than where it meets the half-KT line, giving a deconfinement phase transition directly from the superfluid to the disordered phase. 

An issue left unresolved by the analysis of \cite{Shi:2011} is how the deconfining Ising critical line connects with the usual KT line bordering the phase. Both are transitions out of the superfluid phase into the disordered phase, and traditional lore indicates that a first-order transition must intervene between the two. However, no evidence for a first-order line exists save the already-violated Landau paradigm. A further question is how the Ising deconfinement phase transition relates to the deconfinement transition we found in the flat phase.

We give in this section substantial evidence that the most elegant answer to these questions holds. The KT transition, the deconfinement phase transition, and the deconfinement transition in the flat phase meet at a single multicritical point. Thus the KT line separates the superfluid phase from the usual XY disordered phase, whereas the Ising deconfinement phase transition separates the superfluid phase from the pair-disordered phase.

\begin{figure}[t!]
\begin{center}
\includegraphics[width=0.7\linewidth]{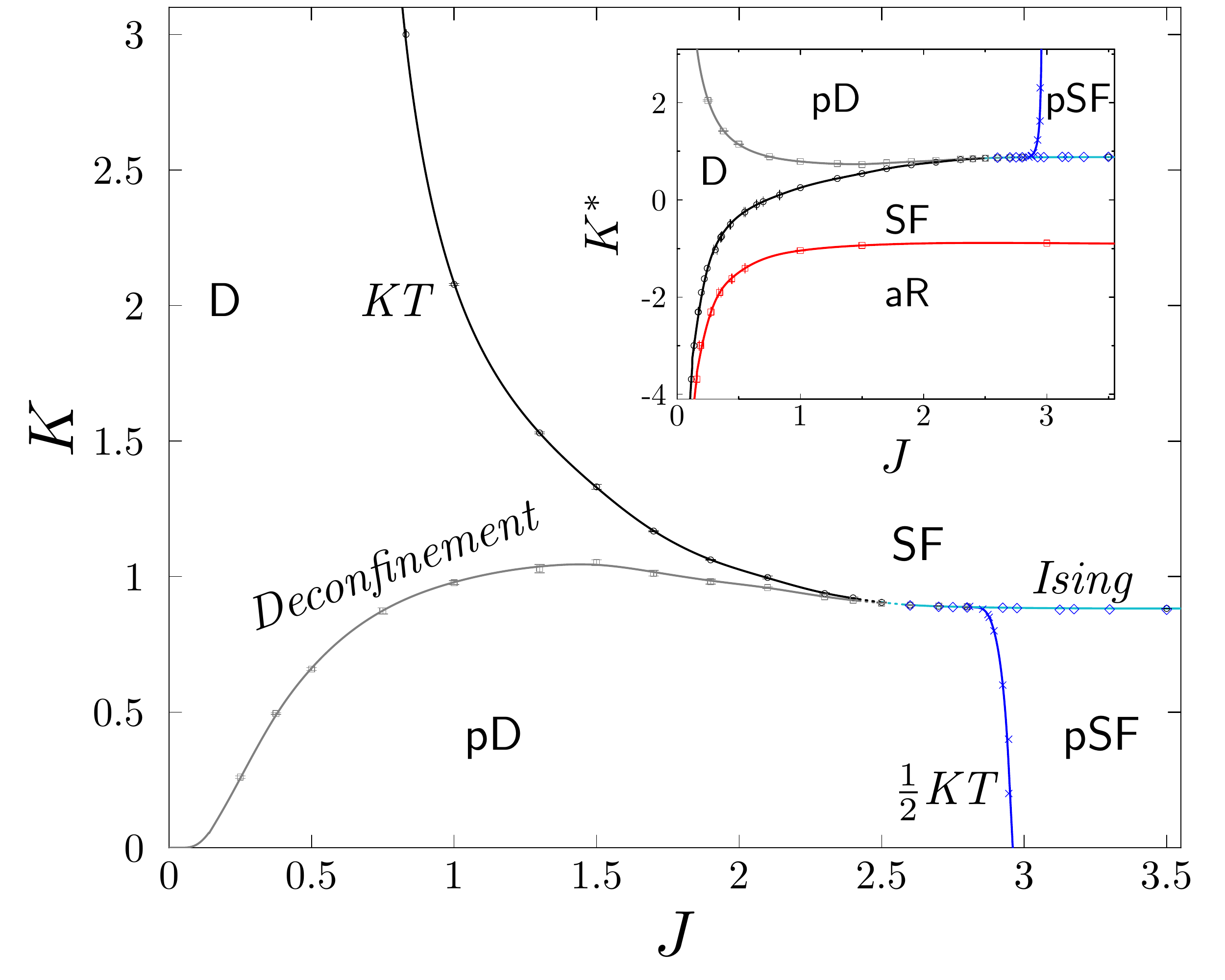}
 \caption{Phase diagram determined from Monte Carlo simulations of the height model (\ref{Zh}). Points are data, and lines are spline fits (except for deconfinement transition near $(J,K)$=(0,0), which comes from (\ref{deconf2})). Data in black, grey and red are from the present work; data in blue (for large $J$) are from Ref.~\cite{Shi:2011}. Main panel: behaviour as a function of $J$ and $K$. Inset: results re-plotted as a function of $K_*$ and extended to negative values of this parameter. }
\label{PhaseDiagram}
\end{center}
\end{figure}

We find by an analysis of the RG equations and
by extensive Monte Carlo simulations the phase diagram shown in Fig.~\ref{PhaseDiagram}. 
As discussed in section \ref{sec:phases}, for $J, K \geq 0$ it contains a superfluid (SF) or rough phase of the height model, a pair superfluid (pSF) phase where the heights are rough but with long-range order in their parity, and a disordered phase where the height field is flat. This last phase is subdivided according to the behaviour of height steps in a sector with non-zero winding number: steps form bound pairs in the pair-disordered (pD) phase  and are unbound on the other side of the deconfinement transition in the usual XY disordered(D) phase. 
Extending this phase diagram to negative values of $K_*$ (see inset to Fig.~\ref{PhaseDiagram}), the rough (SF) phase continues smoothly from small positive $K_*$ (large positive $K$) to small negative $K_*$ but gives way at large negative $K_*$ to a rough (aR) phase with antiferromagnetic long-range order in the parity of the height field.
This diagram matches with precision the above results in the various limits, as well as the analysis of \cite{Shi:2011} for the deconfinement phase transition in the Ising universality class. 

\subsection{Renormalization group analysis}

As a preliminary, we review and extend the renormalization group analysis of the phase diagram, discussed previously in \cite{Shi:2011} and \cite{Zaletel}. The appropriate coupling constants are the stiffness $J_*$ of the bosonic field, the deviation from criticality $\kappa$ of the Ising disorder field, the coupling $z_1$ between these two fields, and the coefficient $z_2$ of a cosine interaction for the bosonic field. To second order in $\kappa$, $z_1$ and $z_2$, the RG equations are \cite{Shi:2011} 
\begin{align}\label{RG}
\frac{{\rm d}z_1}{{\rm d} \ell} = \left( \frac{15}{8} - \frac{\pi}{4J_*}\right) z_1 - \frac{\kappa z_1}{2},\qquad
\frac{{\rm d}J_*}{{\rm d}\ell} = \frac{\pi^2 z_1^2}{4} + Az_2^2, \qquad
\frac{{\rm d} \kappa}{{\rm d} \ell} = \kappa - \frac{z_1^2}{4}
\end{align}
and
\begin{align}
\frac{{\rm d} z_2}{{\rm d} \ell} = \left(2 - \frac{\pi}{J_*} \right)z_2\,,\nonumber
\end{align}
where $\ell$ is the logarithm of the cut-off length and $A$ is a positive, non-universal constant.  

\begin{figure}[htb!]
\begin{center}
\includegraphics[width=0.4\linewidth]{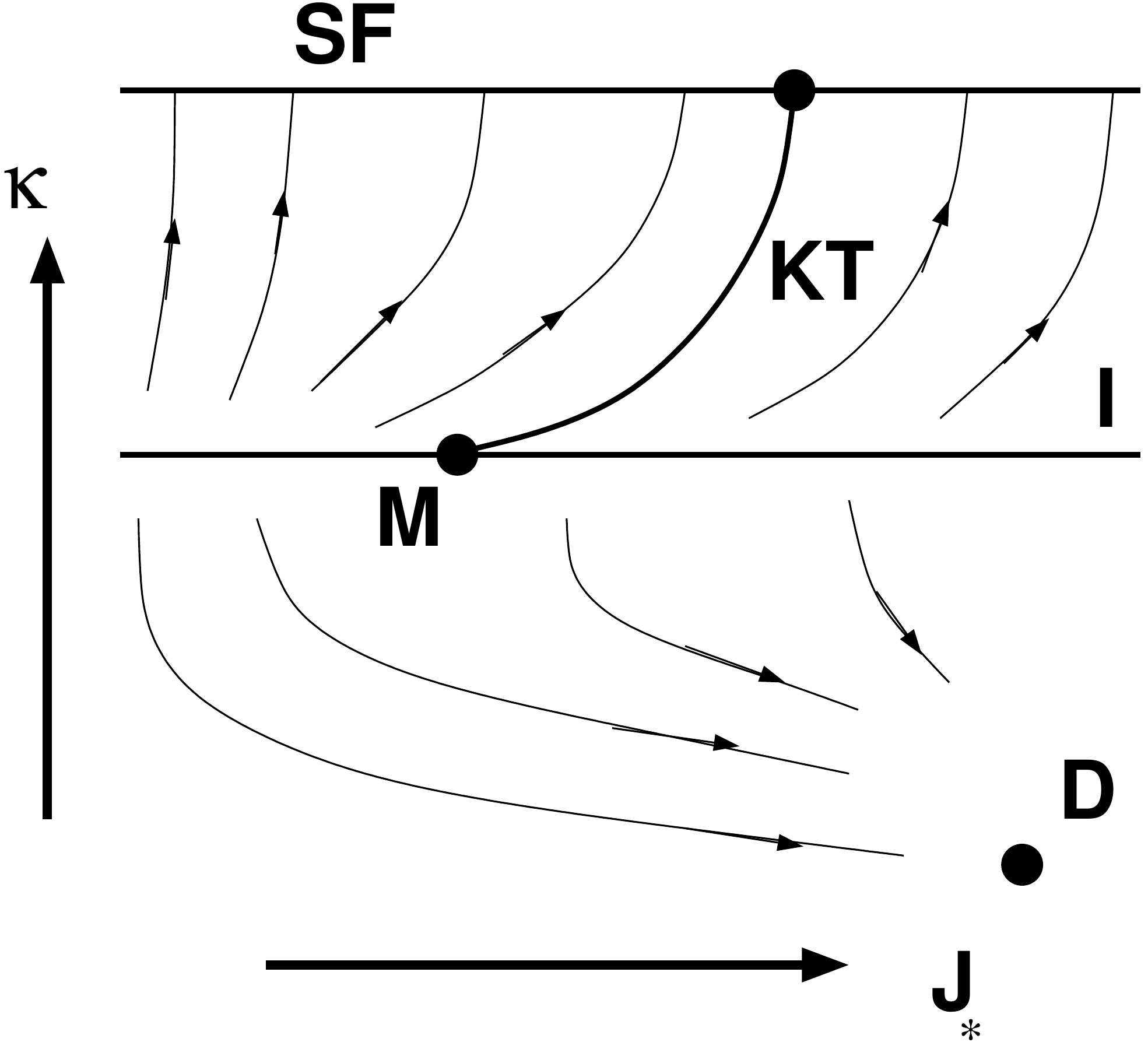}
 \caption{Schematic view of RG flow resulting from (\ref{RG}) for $z_2=0$, projected from $z_1>0$ into the $J_*$--$\kappa$ plane. The KT and Ising separatrices are denoted by KT and I respectively. The multicritical point and the disordered-phase fixed point are indicated by M and D.}
 \label{RGfig}
\end{center}
\end{figure}

To understand the RG flow implied by (\ref{RG}) it is useful to consider first the subspace $z_2=0$, since $z_2$ is RG-irrelevant unless $J_*$ is large ($J_*> \pi/2$).  A schematic view of a section of this flow is shown in Fig.~\ref{RGfig}.
In this subspace, the RG equations have a stable fixed point with $\kappa \to - \infty$ and $z_1$, $J_* \to \infty$ representing the disordered phase. They also have a stable line of fixed points with $z_1=0$, $\kappa \to \infty$ and variable $J_*$. This stable line represents the superfluid phase provided $J_*<\pi/2$, because in that case it is also stable to non-zero $z_2$. The superfluid terminates at $J_*=\pi/2$, since beyond this point $z_2$ is relevant and flow for non-zero $z_2$ is to the disordered phase with $J_* \to \infty$. 

Two separatrices define key features of the RG flow in the $z_2=0$ subspace. In particular, they suggest the existence of a multicritical point where the KT line meets the Ising line with no first-order line intervening. The shapes of the separatrices cannot be determined analytically, but their topology is apparent and can be checked by numerical integration of (\ref{RG}). The ``KT'' separatrix characterises flows at positive $\kappa$ and terminates on the stable line of fixed points. It separates flows that end at points on the superfluid line with $J_*<\pi/2$ from those that end at points with $J_*>\pi/2$.  The ``Ising'' separatrix divides flows that end at the disordered-phase fixed point from those that end at the superfluid-phase fixed line. It contains the line on which $\kappa=z_1=0$. The portion of this line with $J_*<2\pi/15$ is an Ising critical line: only the $\kappa$-direction is unstable but the coupling $z_1$ is a dangerously irrelevant variable \cite{Shi:2011}. Flow lines on the Ising separatrix itself divide into two types: those starting at sufficiently small values of $J_*$ end on the Ising line, while those starting at larger values of $J_*$ run towards $J_* \to \infty$. The locus of the multicritical point is the curve on the Ising separatrix that divides these two types of flow. It is also the curve along which the two separatrices meet.

The description of the transition between the disordered phase and the pair superfluid is straightforward to understand in terms of the bosonic and Ising fields. It occurs for large negative $\kappa$ where the Ising field has spontaneous order. In this regime (\ref{RG}) no longer applies; instead $z_1$ is relevant if $J_* > \pi/8$ and irrelevant otherwise. We do not show this transition in Fig.~\ref{RGfig}.

\subsection{ Simulations}
\label{simulations}

We use Monte Carlo simulations of the height model defined in (\ref{Zh}) to examine the behaviour of the system beyond the limiting regimes of parameter space that can be studied analytically. For most of these simulations we employ a worm algorithm \cite{PS,Alet2003a,Alet2003b} with both single and double worms, as in \cite{Shi:2011}. This algorithm explores all winding-number sectors of the height model. We also require results within specific winding-number sectors that contribute only with small weight to the partition function, and for these we use local updates of height configurations \cite{Wallin1994}.

We locate the positions of the main phase boundaries in this phase diagram -- the transition between the superfluid and disordered phases, and the deconfinement transition -- by studying a Binder cumulant constructed from the Ising variable. This is defined as
\begin{align}\label{binder}
B = 1 - \frac{\langle M^4 \rangle}{3\langle M^2\rangle^2} \quad \mbox{with} \quad M = \sum_a (-1)^{h_a}\,.
\end{align}
Here $\langle \ldots \rangle$ denotes an average over configurations. To identify the superfluid-disordered transition, this average should include all winding-number sectors, but below we locate the deconfinement transition by restricting to a sector with a non-zero even winding number in one direction. 

\begin{figure}[htb!]
\begin{center}
\includegraphics[width=0.45\linewidth]{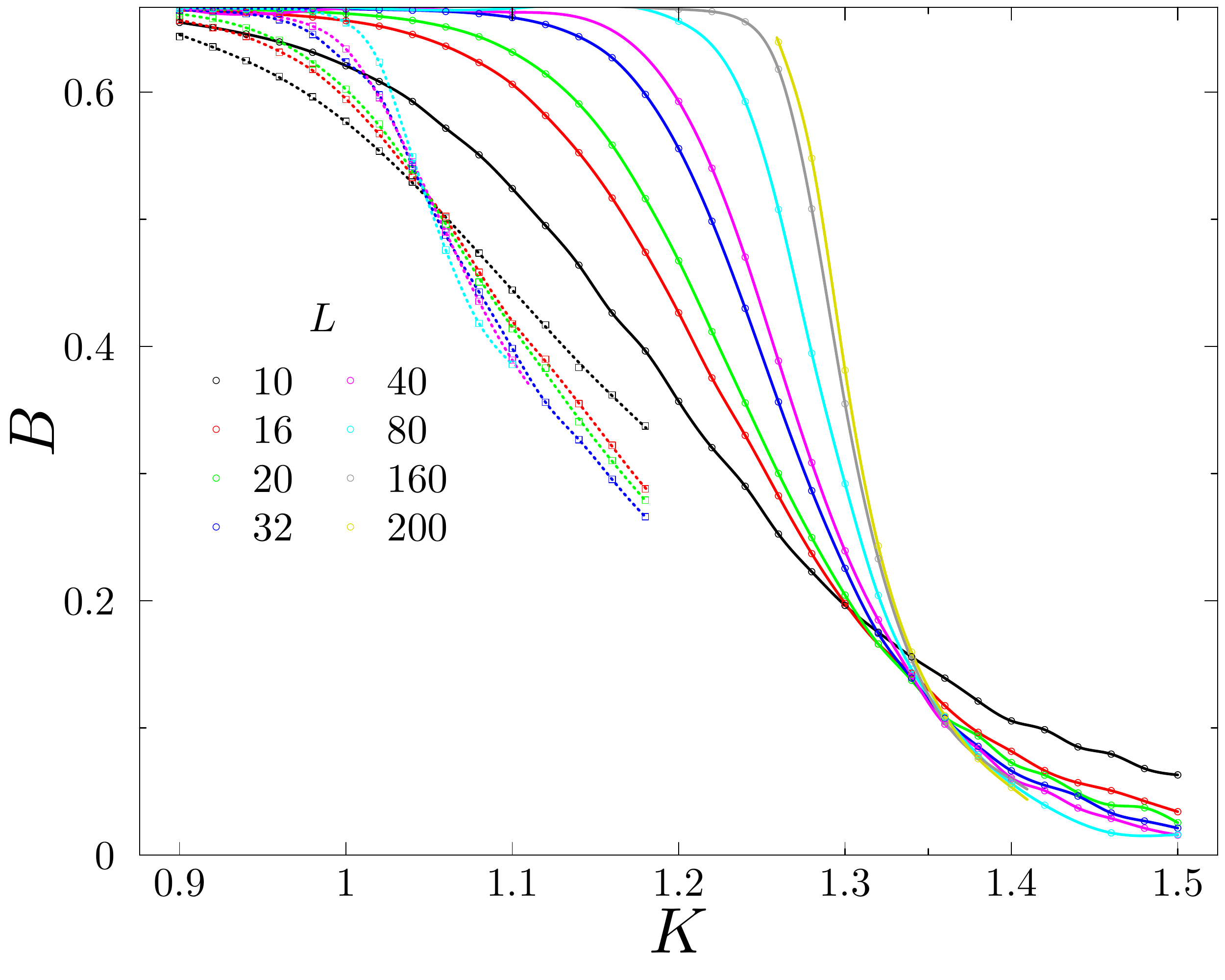}
\includegraphics[width=0.45\linewidth]{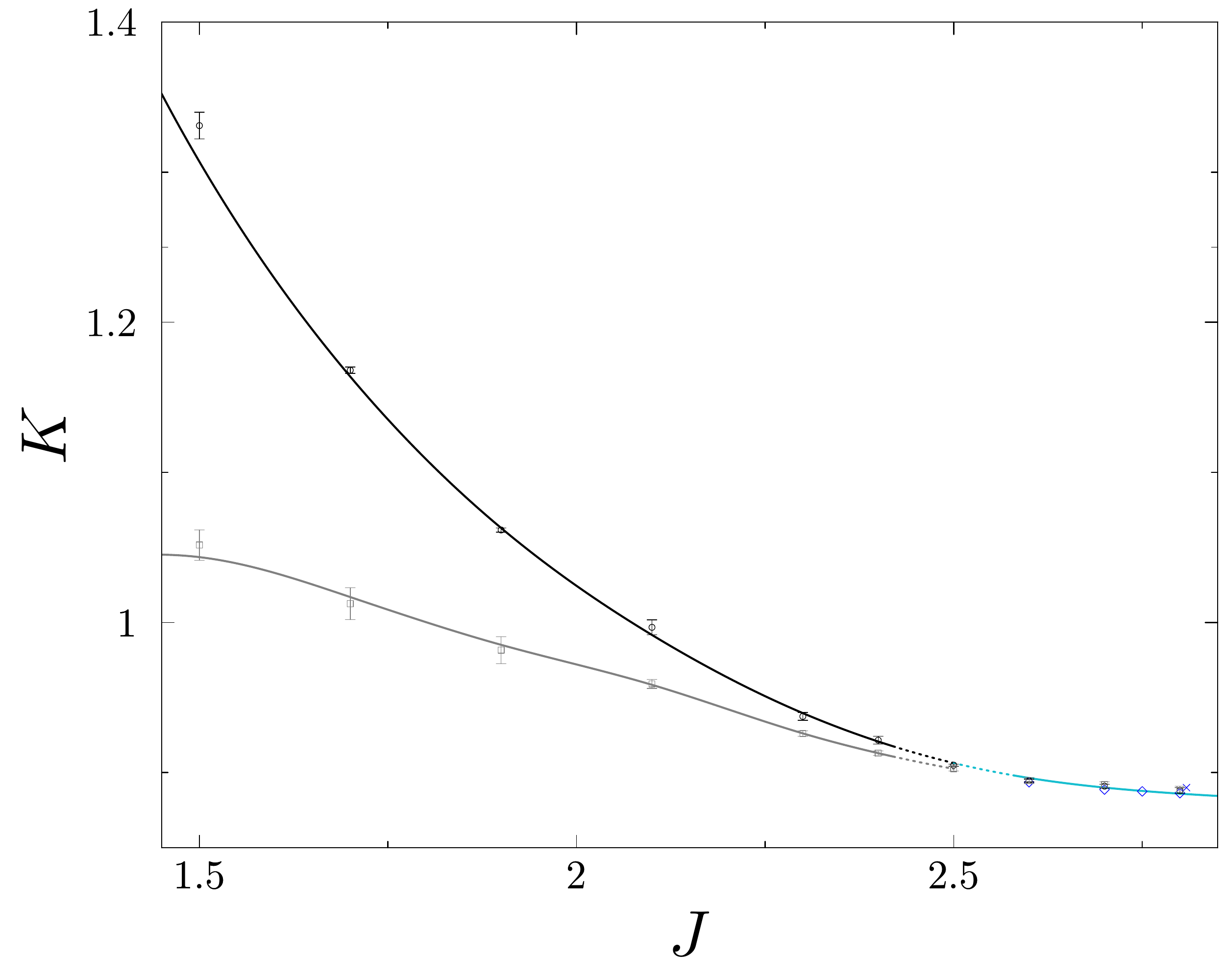}
 \caption{Left panel: evidence for two distinct transitions at $J=1.5$ from the magnetisation Binder cumulant $B$ vs $K$. Circles and solid lines: data averaged over all winding-number sectors, giving the location of the superfluid-disordered phase transition. Squares and dashed lines: data restricted to the winding-number sector $(W_x,W_y)=(0,2)$, giving the location of the deconfinement transition between the disordered and pair-disordered phases.\break Right panel: The ensuing phase diagram near multicritical point.}\label{Binder}
\end{center}
\end{figure}

In the superfluid phase, where height configurations are rough, $M$ has an almost Gaussian distribution with zero mean, so that $\langle M^4 \rangle \approx 3 \langle M^2 \rangle^2$ and the Binder cumulant converges to $B=0$  with increasing system size. By contrast, in the disordered phase of the XY model the height field is pinned and in the dominant winding number sector, $(W_x,W_y) = (0,0)$, heights have only local fluctuations around a global value with definite parity. In consequence, $M$ has a bi-modal distribution concentrated around values $M = \pm L^2 m$, where $m$ is a dual order parameter. The Binder cumulant then converges with system size to the value $B=2/3$. Following a path in $J, K$ that crosses the superfluid-disordered phase boundary, $B$ interpolates between these limiting values, and does so more sharply with increasing system size. As shown in Fig.~\ref{Binder}, the crossing points of curves for successive system sizes converge, giving an estimate of the critical point. We have also determined the position of the critical point using other quantities (the value of $\langle W_x^2\rangle$ and the value of correlation function exponent $\eta$) with consistent results.

To locate the deconfinement transition within the disordered phase, we evaluate $B$ in the winding number sector $(W_x,W_y) = (0,2)$. If the height steps imposed by boundary conditions in this sector are bound in pairs, they do not affect the distribution of $M$ or the value of $B$. Conversely, if these steps are unbound, typical configurations consist of two domains with opposite values for the parity of the height field. These domains are approximately rectangular, because the steps that separate them have transverse fluctuations of amplitude much smaller than sample size. Variations in the width of these domains give $M$ a single-peaked distribution centred on zero, and so in this sector the Binder cumulant converges to small values both in the superfluid phase and in the disordered phase with unbound steps.  It is clear from the data in Fig.~\ref{Binder} that the step-binding transition occurs at a point distinct from the superfluid-disordered phase boundary.

\begin{figure}[htb!]
\begin{center}
\includegraphics[width=0.55\linewidth]{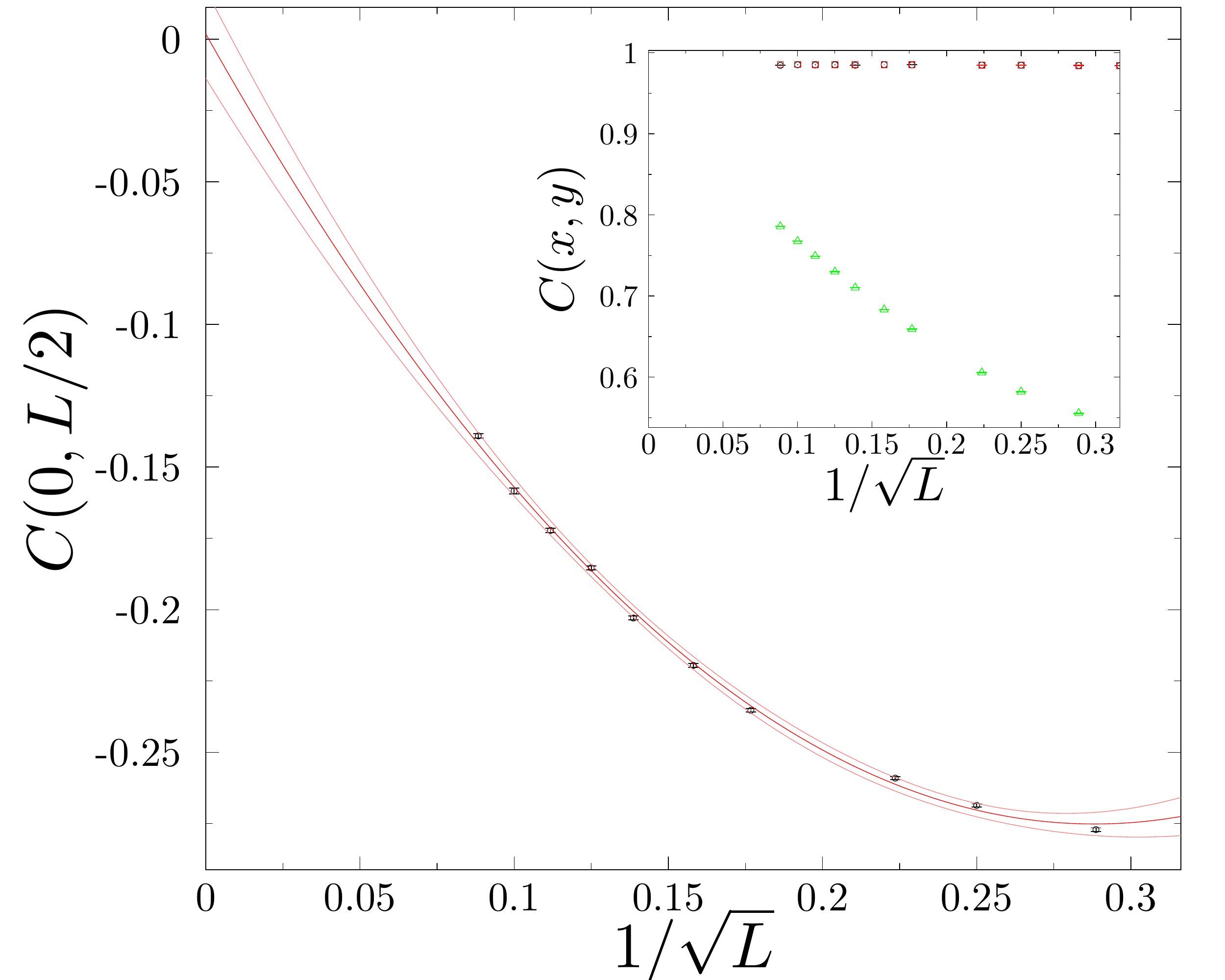}
 \caption{Behaviour of $C(x,y)\equiv \langle (-1)^{h(x,y) - h(0,0)}\rangle$ in the disordered phase, studied in the winding number sector $(W_x,W_y)=(0,2)$. Main panel: behaviour for $J=0.5$, $K=2$, when steps are unbound. Decay with increasing system size $L$ of $C(0,L/2)$  transverse to average step direction. Points: data; lines: fits to $C(0,L/2) = a+bL^{-1/2}+ c L^{-1}$ together with the statistical 95\% confidence interval. Inset: $(i)$ absence of decay parallel to average step direction (green points), from $C(L/2,0)$  for $J=0.5$, $K=2$; $(ii)$ absence of decay in either the perpendicular direction ($C(0,L/2)$, black points) or the parallel direction ($C(L/2,0)$, red points) for $J=2$, $K=0.5$, when steps are bound.}\label{Ising}
\end{center}
\bigskip
\end{figure}

A probe of domain geometry is provided by the behaviour of the Ising-spin two-point correlator 
\begin{align}
C(x,y) = \langle (-1)^{h(x,y) - h(0,0)}\rangle\,.
\end{align}
of height parities in the winding number sector $(W_x,W_y) = (0,2)$. When steps are bound in pairs, this correlator is almost isotropic and has a non-zero limiting value at large separation. By contrast, when steps are unbound the correlator is highly anisotropic. In this sector, steps run approximately parallel to the $x$-direction, so the value of $C(0,L/2)$ probes correlations in the direction transverse to the steps. It is positive if the points $(x,y) = (0,0)$ and $(0,L/2)$ are predominantly in the same domain, and negative if they are predominantly in opposite domains. As illustrated in Fig.~\ref{Ising} (main panel), it decreases towards zero with increasing system size. This is exactly as expected if the width of each domain fluctuates over the full range $(0,L)$, so that both alternatives contribute equally. To test correlations in a direction parallel to the steps, we compute $C(L/2,0)$ at the same point in the phase diagram: see Fig.~\ref{Ising} (inset). Without transverse fluctuations of steps, the points $(x,y) = (0,0)$ and $(L/2,0)$ would always lie in the same domain. As transverse fluctuations are expected to be small, we anticipate that this correlator should be non-zero, as is indeed observed.
 \begin{figure}[htb!]
\begin{center}
 \includegraphics[width=0.45\linewidth]{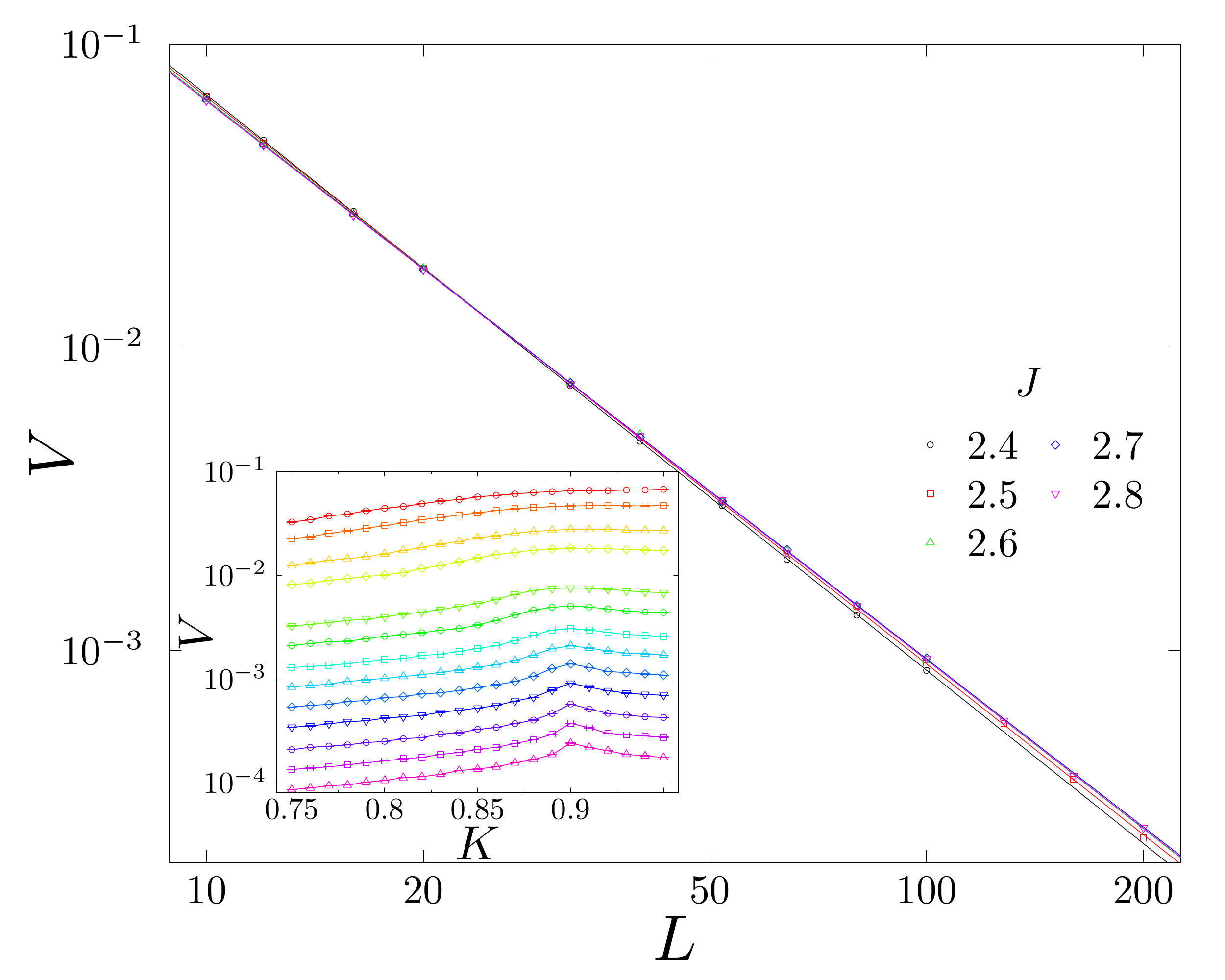}
 \includegraphics[width=0.45\linewidth]{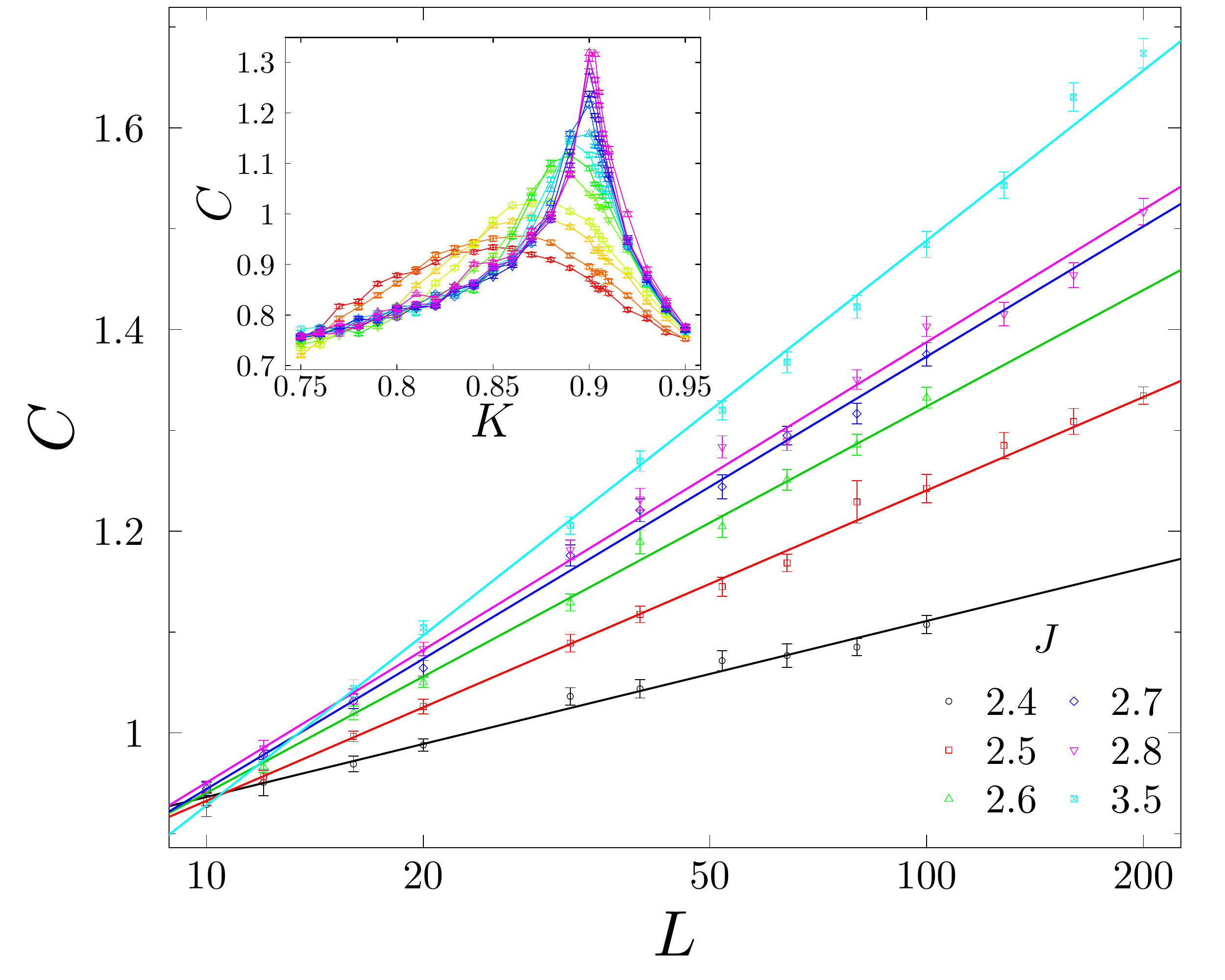}
 \caption{Evidence for continuous transitions from energy fluctuations. Phase boundary is crossed by varying $K$ at the indicated values of $J$. Main panels: left -- dependence of maximum value of energy Binder cumulant $V$ on system size (log-log scale);  right  -- dependence of maximum value of heat capacity $C$ on system size (log-linear scale). A first-order transition would be signalled by finite $V$ at large $L$ and by maximum $C$ $\propto L^2$. Insets: $V$ (left) and $C$ (right) vs $K$ for various $L$ at $J=2.5$.}\label{Continuous}
\end{center}
\end{figure}

An important issue our simulations address is the nature of the phase transition between the superfluid and disordered phases, as a function of position along the phase boundary. Previous work \cite{Shi:2011} yielded the remarkable result that critical behaviour is in the Ising universality class for a portion of the boundary close to the pair superfluid phase ($J > 2.6$) but left open the character of the transition for smaller values of $J$, where the critical value of $K$ is larger. At sufficiently large $K$ a conventional KT transition is expected, as described in section \ref{sec:phases}. At intermediate $K$, however, Landau arguments suggests a first-order transition. Our data shown in Fig.~\ref{Continuous} (left panel) provide strong evidence that the transition is continuous all along this phase boundary, with no sign of a first-order transition. We use a standard probe of the order of a transition, a Binder cumulant $V$ computed from fluctuations in the energy $E$ of the system, defined by
\begin{align}
V = \frac{\langle E^4 \rangle}{\langle E^2 \rangle^2} - 1\,.
\end{align}
A first-order transition with a latent heat is signalled by a finite limiting value for $V$ at large system size \cite{Binder1986}. Instead we observe power-law decrease of $V$ with $L$ at all points studied on the superfluid-disordered phase boundary, indicating exclusively continuous transitions. An alternative test of the order of the transition is provided by the size-dependence of the maximum in the heat capacity: if there is a latent heat at the transition, this maximum is proportional to $L^2$; at a continuous transition in the Ising universality class it is proportional to $\ln L$; and at a KT transition it has a finite limiting value for large $L$. The data presented in Fig.~\ref{Continuous} (right panel) give no suggestion of a first-order transition, but are consistent with Ising critical behaviour for $J\geq 2.6$.

\begin{figure}[htb!]
\begin{center}
\includegraphics[width=0.54\linewidth]{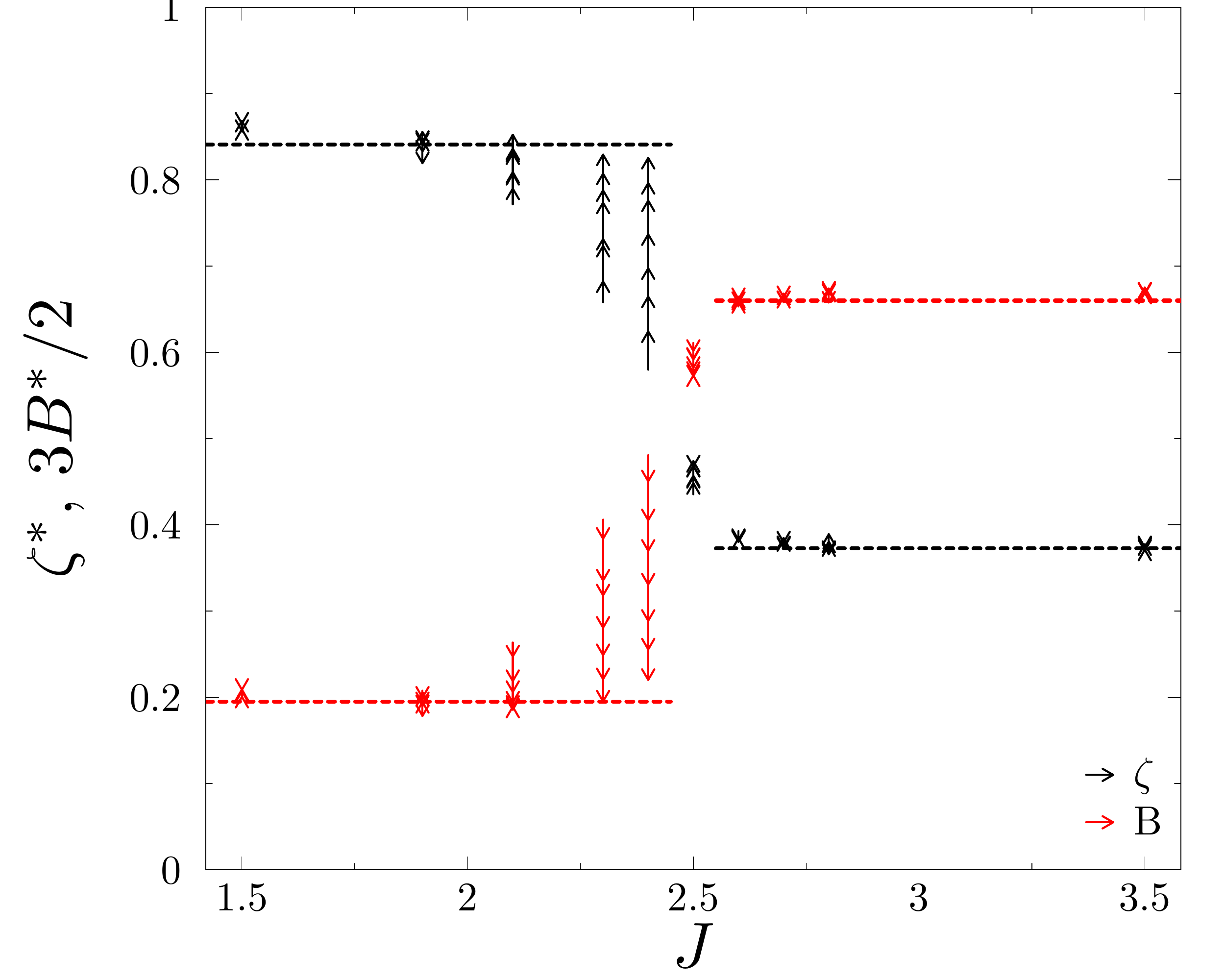}
 \caption{Universality classes at transition from superfluid phase: values of $B$ and $\zeta$ at points on the phase boundary, shown for system sizes $32\leq L \leq 160$ with arrows giving flow in direction of increasing system size. $K$ is tuned to the critical value for each $J$. Dashed horizontal lines indicate critical values of  $B$ and $\zeta$ at the Ising transition ($J > 2.5$) and the KT transition ($J<2.5$): for $B$ these are estimated using simulations far from $J_c$, and for $\zeta$ they are theoretical values from \cite{Shi:2011}.}\label{UniversalityClass}
\end{center}
\end{figure}

To identify the critical behaviour at points on the superfluid-disordered phase boundary, we examine the behaviour of two quantities as a function of position along the phase boundary. One of these is the Binder cumulant $B$, defined in (\ref{binder}); the other is a partition function ratio
\begin{align}
\zeta = \frac{Z_{W_x\, {\rm even}, W_y\,{\rm odd}} + Z_{W_x\, {\rm odd},W_y\,{\rm even}} }{2Z_{W_x\, {\rm even},W_y\,{\rm even}} }\,,
\end{align}
introduced in \cite{Shi:2011}. Both $B$ and $\zeta$ are expected to be universal at the critical point, with values that are characteristic of the universality class. Since finite size effects are significant over a large portion of the phase boundary, we examine dependence of $B$ and $\zeta$ on system size. Fig.~\ref{UniversalityClass} provides clear evidence that the critical point is in the Ising universality class for $J >J_c$, and in the KT universality class for $J < J_c$, with $J_c \approx 2.5$.

These Monte Carlo simulations provide extensive support for the picture of the phase structure of the generalised XY model developed in Secs.~\ref{sec:phases} and \ref{sec:deconfinement}, and summarised in  figure \ref{PhaseDiagram}. In particular, we believe there is convincing evidence of two types of disordered phase, separated by a deconfinement transition. Moreover, this transition meets the Ising deconfinement phase transition and the KT transition at a single multicritical point. This meeting ensures that the latter two govern transitions from the superfluid phase to the pair-disordered phase and the conventional XY disordered phase respectively.

\bigskip

We close by noting that there are several interesting generalisations of the model considered here. One can modify the potential to instead allow a vortex to fractionalise into $q$ parts. Extensive Monte Carlo simulations have been done for the cases $q=3$ and $8$ \cite{Arenzon1,Arenzon2,Arenzon3}, and found a nematic phase analogous to the pair-superfluid phase in the $q=2$ case described here.  It thus would be very interesting to understand if either of the deconfinement transitions occurs for higher $q$. Similar physics also can occur when generalising the models to three dimensions \cite{Arenzon3}. Moreover, transitions governing deconfinement of half-vortices do occur in three dimensions \cite{Leo1,Leo2}. Since the deconfined quantum criticality scenario \cite{Senthil} was originally developed for 2d quantum theories, it would be quite exciting if any of the new results described here applied to 3d classical theories. 

\bigskip\bigskip

We would like to thank Austen Lamacraft for many useful discussions, and Jesper Jacobsen, Mike Zaletel and Joel Moore for sharing their unpublished results. The work of JTC and PF was supported by EPSRC through grant EP/N01930X.

\bibliographystyle{iopart-num}

\bibliography{deconfinement}

\end{document}